\documentclass[10pt,conference]{IEEEtran}
\IEEEoverridecommandlockouts
\usepackage{graphics}
\usepackage[update,prepend]{epstopdf}
\usepackage[pdftex]{graphicx}
\usepackage{stfloats}
\usepackage{array}
\usepackage{float}
\usepackage[cmex10]{amsmath}
\usepackage{amsmath,amsfonts,amssymb}
\usepackage{graphicx,graphics}
\usepackage{enumerate}
\usepackage{textcomp}
\usepackage{color}
\usepackage{verbatim}
\usepackage{amsthm}
\usepackage{multirow}
\usepackage{subcaption}
\usepackage{siunitx}
\usepackage{amsthm}
\usepackage{cleveref}
\usepackage{cite}
\usepackage{pdfpages}
\usepackage[font=footnotesize]{caption}
\usepackage{fixltx2e}
\usepackage[USenglish]{babel}
\usepackage{algorithm}
\usepackage[noend]{algpseudocode}

\usepackage{geometry}
\newcommand{\e}{\mathrm{e}}

\def\BibTeX{{\rm B\kern-.05em{\sc i\kern-.025em b}\kern-.08em
    T\kern-.1667em\lower.7ex\hbox{E}\kern-.125emX}}
\begin{document}

\title{Reinforcement Learning for Mitigating Intermittent Interference in Terahertz Communication Networks
\thanks{This research was partially supported by the U.S. National Science Foundation under Grant CNS-1941348.}
}
\author{\IEEEauthorblockN{Reza Barazideh}
\IEEEauthorblockA{\textit{Department of ECE} \\
\textit{Kansas State University}\\
 Manhattan, KS, USA \\
rezabarazideh@ksu.edu}
\and
\IEEEauthorblockN{Omid Semiari}
\IEEEauthorblockA{\textit{Department of ECE} \\
\textit{\hspace{0.19cm}University of Colorado Colorado Spring}\\
Colorado Springs, CO, USA \\
osemiari@uccs.edu}
\and
\IEEEauthorblockN{Solmaz Niknam}
\IEEEauthorblockA{\textit{Department of ECE} \\
\textit{Virginia Tech}\\
Blacksburg, VA, USA \\
slmzniknam@vt.edu}
\and
\IEEEauthorblockN{Balasubramaniam Natarajan}
\IEEEauthorblockA{\textit{Department of ECE} \\
\textit{Kansas State University}\\
 Manhattan, KS, USA \\
bala@ksu.edu}
\vspace{-0.5cm}
}

\maketitle
\begin{abstract}
Emerging wireless services with extremely high data rate requirements, such as real-time extended reality applications, mandate novel solutions to further increase the capacity of future wireless networks. In this regard, leveraging large available bandwidth at terahertz frequency bands is seen as a key enabler. To overcome the large propagation loss at these very high frequencies, it is inevitable to manage transmissions over highly directional links. However, uncoordinated directional transmissions by a large number of users can cause substantial interference in terahertz networks. While such interference will be received over short random time intervals, the received power can be large. In this work, a new framework based on reinforcement learning is proposed that uses an adaptive multi-thresholding strategy to efficiently detect and mitigate the intermittent interference from directional links in the time domain. To find the optimal thresholds, the problem is formulated as a multidimensional multi-armed bandit system. Then, an algorithm is proposed that allows the receiver to learn the optimal thresholds with very low complexity. Another key advantage of the proposed approach is that it does not rely on any prior knowledge about the interference statistics, and hence, it is suitable for interference mitigation in dynamic scenarios. Simulation results confirm the superior bit-error-rate performance of the proposed method compared with two traditional time-domain interference mitigation approaches.

\vspace{-.5em}
\end{abstract}



\section{Introduction}
Despite major advancements in fifth-generation (5G) systems, new solutions are still required to increase the capacity of wireless networks and handle the continuous exponential growth in
mobile data traffic. In particular, communications at high frequencies above the conventional sub-6 GHz bands is seen as a key enabler which allows leveraging the large available bandwidth and achieving very high data rates~\cite{Omid_TWC_2017,Omid_WC_2019}. Nonetheless, emergence of new technologies such as wireless extended reality (XR), connected and autonomous vehicles, and factory automation will introduce new challenges for future wireless networks beyond the 5G system. In fact, to support extremely high data rates needed for such real-time applications, new wireless solutions must be developed that enable exploiting the large available bandwidth at very high frequencies (above 100 GHz)~\cite{Rappaport_TeraHz_2019}.

\begin{figure}[t]
\centering
\includegraphics[width=.35\textwidth,height=46mm]{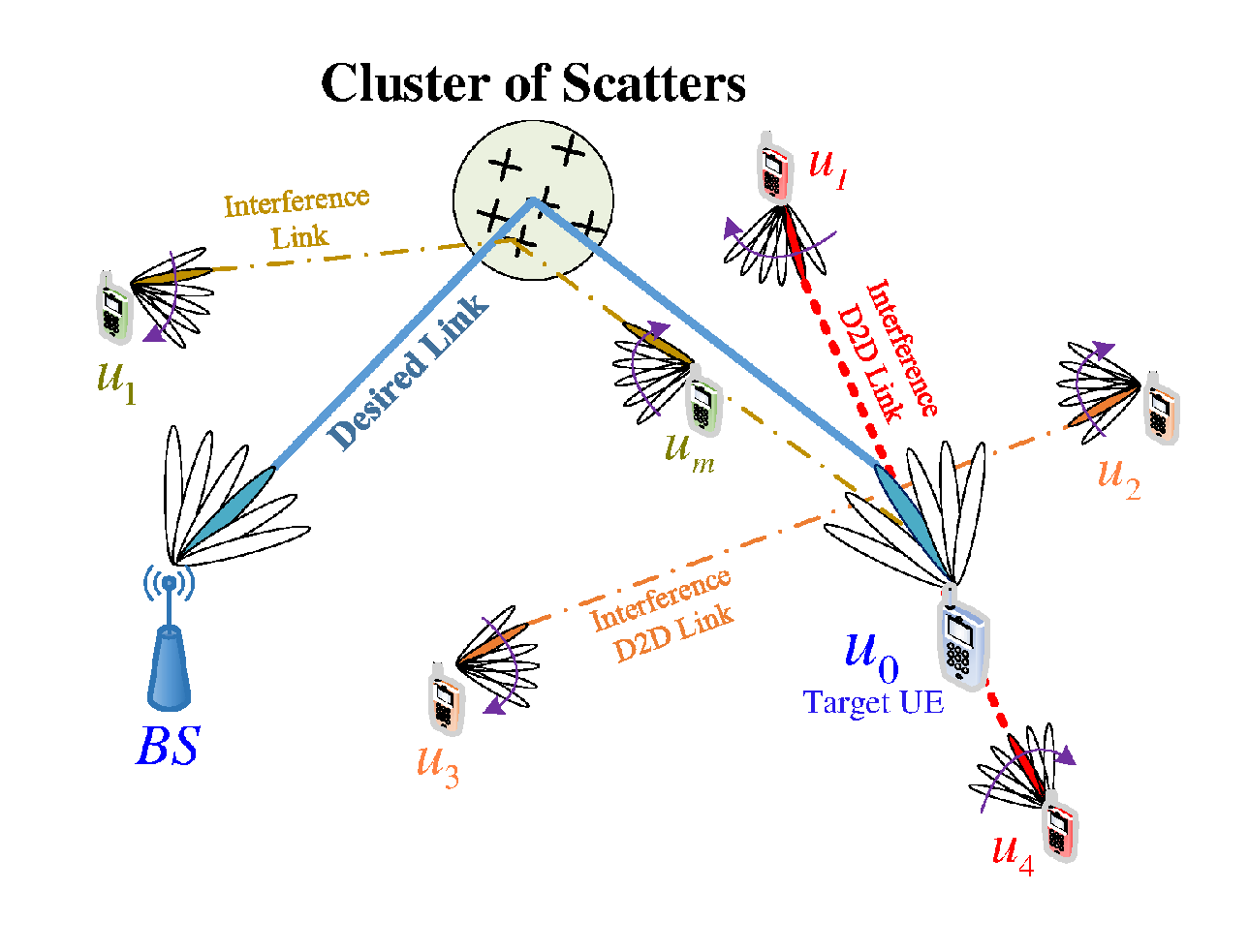}
\caption{An example scenario for uncoordinated directional THz links, creating intermittent interference at the target UE's receiver.}
\label{fig:Interference_Diagram}
\vspace{-.5cm}
\end{figure}

In addition to the substantial available bandwidth, particularly at the terahertz (THz) frequency range (commonly referred to as the frequencies within 0.1-10 THz~\cite{Rogalski_Terahertz_2011}), communications over very high frequencies allow deployment of small-size antenna arrays with many antenna elements. Hence, despite the high atmospheric propagation loss at THz bands, the communication range can be extended by leveraging large array gains over highly directional THz links. However, such pencil-beams can cause a significantly large interference, if the receiver's beam is accidentally directed toward a dominant multi-path of an interference link. In addition, transceivers that operate at high frequencies need to constantly change the direction of their beams (i.e., perform beam training) due to mobility of users or changes in the propagation environment. Therefore, the interference from directional links is typically intermittent and occur at random time intervals~\cite{Cross2018niknam,Regime2018niknam}. Moreover, the interference power can be large and even exceed the received power over the desired link.  An example scenario is shown in Fig.~\ref{fig:Interference_Diagram} in which directional transmissions from interfering users cause intermittent interference at the target receiver $u_0$.
To mitigate the inevitable strong interference from directional links, space-domain interference reduction methods use antenna arrays with adaptive beamforming to suppress the interfering signals by steering the beam to different directions or by placing nulls in the antenna gain pattern toward the direction of the interfering signals~\cite{Joham_MIMO_ZF_2005}. This strategy requires complete knowledge of the position of interferer, channel, and signal identity~\cite{Wiesel_ZFBF_2008}. However, in realistic scenarios, such information is not typically available at the receiver. Therefore, new techniques are required to suppress the inter-beam interference in dynamic scenarios without relying on any prior knowledge about the interference links.

A plethora of techniques have been studied in the literature aiming at alleviating the impacts of intermittent interference~\cite{Khodam_TVT,MultiThershold_Rozic_2018,AdaptiveNoiseMitigation-2010}. The high amplitude and the short duration of the interference can be exploited to detect and mitigate the effect of interference. Conventional threshold-based nonlinear preprocessors such as clipping, blanking, and their combination fall in this category~\cite{MultiThershold_Rozic_2018}. Adaptive analog nonlinear filtering in high acquisition bandwidth is used in~\cite{Khodam_TVT} to determine the threshold in orthogonal frequency-division multiplexing (OFDM) signals. An optimization problem is formulated in~\cite{AdaptiveNoiseMitigation-2010} to find the optimum threshold. Recently, by exploiting emerging machine learning techniques, it has been shown that deep neural network (DNN) can also be used as a powerful tool for interference mitigation~\cite{Khodam_CCWC_ML}.

Although interesting, the performance of all prior threshold-based methods are highly sensitive to the selected thresholds~\cite{MultiThershold_Rozic_2018,Khodam_TVT}. Therefore, finding the optimum threshold is the main challenge for these methods as the threshold must be dynamically determined according to channel variations and model mismatches. The proposed approach in~\cite{AdaptiveNoiseMitigation-2010} is not robust in model mismatch. Moreover, the DNN-based technique in \cite{Khodam_CCWC_ML} requires large labeled dataset for training the model which may not be available in many applications. Furthermore, most of existing works focus on mitigating impulsive noise in sub-6 GHz networks, while limited work exists for managing the intermittent interference from directional links at high millimeter wave and THz frequencies.

The key contribution of this work is a novel interference mitigation framework, based on reinforcement learning (RL), which enables the receiver to effectively reduce the interference power received intermittently from highly directional links in a THz network. The proposed framework develops a multi-threshold clipping strategy to dynamically change the cut-off threshold for reducing the interference power. To this end, a multi-armed bandit (MAB) is designed to determine the effective values of multi-thresholds in memoryless nonlinear preprocessor. The proposed approach provides near-optimum threshold values even in a non-stationary environment. The key advantage of the proposed approach is to learn the optimal threshold values online with a very low complexity, as compared with other learning schemes such as DNN which requires large training datasets. The simulation results show the superiority of our approach, in terms of the bit-error-rate (BER), over conventional techniques without increasing the complexity of the receiver.


\vspace{-.2cm}

\section{System Model}\label{sec:Models}


Consider a THz network shown in Fig.~\ref{fig:Interference_Diagram} consists of a base station (BS), a target user equipment (UE) $u_0$, and a set of interfering UEs in a set $\mathcal U=\{u_1,...,u_{I}\}$ that communicate with one another over direct device-to-device (D2D) links. The UEs and the BS are equipped with uniform linear arrays (ULAs) with isotropic  antennas. The antenna elements are equally spaced by a distance $d=\lambda/2$, where $\lambda$ is the wavelength at the carrier frequency $f_c=140$ GHz.


OFDM is considered as the underlying multi-carrier technique for sending information over the cellular and D2D links. Let, $\textbf{s}=[s_{0},s_{1},...,s_{K-1}]$ represents a frequency domain OFDM symbol with $K$ subcarriers. According to the OFDM modulation, the time domain symbol $\textbf{x}_0=[x_{0,0},x_{0,1},...,x_{0,K-1}]$ is generated by computing the inverse discrete Fourier transform (IDFT) of $\textbf{s}$ as expressed by
\begin{equation}
{\textbf{x}_{0}} = \textbf{F}_K^{\mathcal H}{\textbf{s}},
\end{equation}
where $\mathcal H$ is the Hermitian operator, and $\textbf{F}_K$ denotes the K-point unitary discrete Fourier transform (DFT) matrix. A cyclic prefix (CP) $\textbf{x}_0^{CP}=[x_{0,K-\mu},...,x_{0,K-1}]$ with length $\mu$ is inserted at the beginning of the OFDM symbol $\textbf{x}_0$ to mitigate the inter-symbol-interference (ISI) and simplify the equalizer structure. The constructed time domain OFDM symbol is transmitted through the transmitter antenna array.


\vspace{-.2cm}

\subsection{Channel Model}\label{sec:Channel Model}
Due to the signal propagation characteristics at high frequencies, i.e., poor penetration through objects and reflection from surfaces, we consider a single-path fading channel (for the desired and interference links) as a widely adopted channel model~\cite{Akdeniz_mmWaveChannelModel_2014}. In particular, the channel matrix for the desired link can be modeled as
\begin{equation}\label{eq:CH_Model}
{\textbf{H}} = {\alpha}\textbf{u}({\theta _r})\textbf{v}^{\mathcal{H}}({\theta _t}),
\end{equation}
where $\theta _r,\theta _t \in [-\pi/2, \pi/2]$ are angle of arrival (AoA) and angle of departure (AoD), respectively. Here, $\alpha$ is the random complex fading gain for an arbitrary link and follows i.i.d. Rayleigh distribution. In addition, $\textbf{u}$ and $\textbf{v}$ are receiver and transmitter array response vectors, respectively, given by
\begin{align}\label{eq:Response_Vector}\notag
\textbf{u}({\theta _r}) &= {\left[ {1,{e^{-j\frac{{2\pi {d}}}{\lambda }\sin ({\theta _r})}},...,{e^{-j({N_r^i} - 1)\frac{{2\pi {d}}}{\lambda }\sin ({\theta _r})}}} \right]^T}, \\
\textbf{v}({\theta _t}) &= {\left[ {1,{e^{-j\frac{{2\pi {d}}}{\lambda }\sin ({\theta _t})}},...,{e^{-j({N_t^i} - 1)\frac{{2\pi {d}}}{\lambda }\sin ({\theta _t})}}} \right]^T},
\end{align}
where, $N_r^i$ and $N_t^i$ denote the number of antennas in the $i$-th receiver and transmitter, respectively. Given that AoA and AoD change much slower compared with the fading channel, it is assumed that $\theta _r$ and $\theta _t$ are fixed in the duration of one received OFDM frame~\cite{VA_BeamwidthandChVar_2017}. The channel model for the interference links also follows \eqref{eq:CH_Model} with different AoA and AoD for each link. Considering the proposed channel model and deploying analog beamforming at both the BS and the UEs, the received OFDM symbol in presence of $I$ interferers and the receiver noise is
\begin{align}\label{eq:Recived Signal} \notag
{\textbf{r}} =\sqrt {{P_{b}}} \textbf{w}_{0}^{\mathcal{H}}{\textbf{H}_{0}}{\hat{\textbf{w}}_{b}}{\textbf{x}_{0}} + \sum\limits_{i = 1}^{{I}} \sqrt {{P_{i}}} \textbf{w}_{0}^{\mathcal{H}}{\textbf{H}_{i}}{\hat{\textbf{w}}_{i}}{\textbf{x}_{i}} + {\textbf{n}},
\end{align}
where $P_{b}$ and $P_{i}$ are the average transmit power of the serving BS and $i$-th interfering UE, respectively. The channel matrix of desired and $i$-th interfering link are represented by $\textbf{H}_{0}\in {\mathbb{C}^{N_r^0\times N_t^{b}}}$ and $\textbf{H}_{i}\in \mathbb{C}^{N_r^0\times N_t^i}$, respectively. Here, $N_t^b$ represents the number of transmitter antennas in serving BS. Moreover, $\hat{\textbf{w}}_{b} \in \mathbb{C}^{N_t^{b}}$ is the BS beamforming vector with $\|\hat{\textbf{w}}_{b}\|_2^2=1$, ${\textbf{w}}_{0} \in \mathbb{C}^{N_r^0}$ is the $u_0$ combining vector with $\|{\textbf{w}}_{0}\|_2^2=1$, $\hat{\textbf{w}}_{i} \in \mathbb{C}^{N_t^i}$ is the $i$-th interfering UE beamforming vector with $\|\hat{\textbf{w}}_{i}\|_2^2=1$. $\textbf{n}\sim\mathcal{CN}(\textbf{0}, {N_0}{B}\textbf{I})$ denotes the additive white Gaussian noise (AWGN), with one-sided power spectral density $N_0$, $B$ is the system bandwidth, and $\textbf{I}$ denotes the identity matrix.

\subsection{Interference Model}\label{sec:Interference Model}

\begin{figure}
\centering
\includegraphics[width=.40\textwidth,height=31mm]{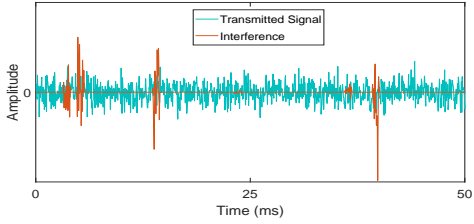}
\caption{One realization of received intermittent interference.}
\label{fig:Interference}
\vspace{-.3cm}
\end{figure}

To establish directional THz links, each pair of D2D users must sweep their beams (by selecting different beamforming weights) to find the best spatial direction that yields the maximum received power. The frequency of beam sweep depends on the size of beamforming codebook which is generally determined from the required resolution of the angular search. As depicted in Fig.~\ref{fig:Interference_Diagram}, during the beam-training phase, the AoA of the interference signals could be exactly or almost aligned with the AoA of the desired signal at the input of the target receiver $u_o$. The power of the interference mostly depends on how the AoA of the desired link and interference links are aligned. As the interfering UEs sweep their beams by selecting different beamforming vectors from their codebook, the received interference at the target UE will appear at random time instances and with random power, as shown in Fig.~\ref{fig:Interference}. In fact, the uncoordinated D2D transmissions during the beam training phase results in shot-like, intermittent interference at the desired link and the power of interference signals could vary significantly relative to the power of the desired signal.

To mitigate this interference in the time domain, we adopt a proper nonlinear function as
\begin{equation}\label{eq:received signal_time}
{{{\mathbf{\hat{r}}}}} = g({{\mathbf{r}}},{{\mathbf{a}}}),
\end{equation}
where ${{{\mathbf{\hat r}}}}$ represents the received signal after interference suppression, $g(.)$ is the nonlinear function, and vector $\textbf{a}$ contains the set of selected pair thresholds-levels. More details about $g(.)$ and vector $\textbf{a}$ are provided in sections \ref{sec:Problem} and \ref{sec:RL}, respectively. The frequency domain representation of the received signal can be obtained by using DFT as
\begin{equation}\label{eq:received signal_Freq}
{{\mathbf{y}}} = {{\mathbf{F}}_K}{{{\mathbf{\hat r}}}}.
\end{equation}
We note that the key advantage of the time-domain interference mitigation is to reduce the interference power prior to performing DFT at the OFDM receiver. Otherwise, large interference would spread over all subcarriers after performing DFT which can severely impact the performance of the receiver.

\vspace{-.15cm}

\section{Problem Formulation}\label{sec:Problem}

\begin{figure}
\centering
\includegraphics[width=.33\textwidth,height=32mm]{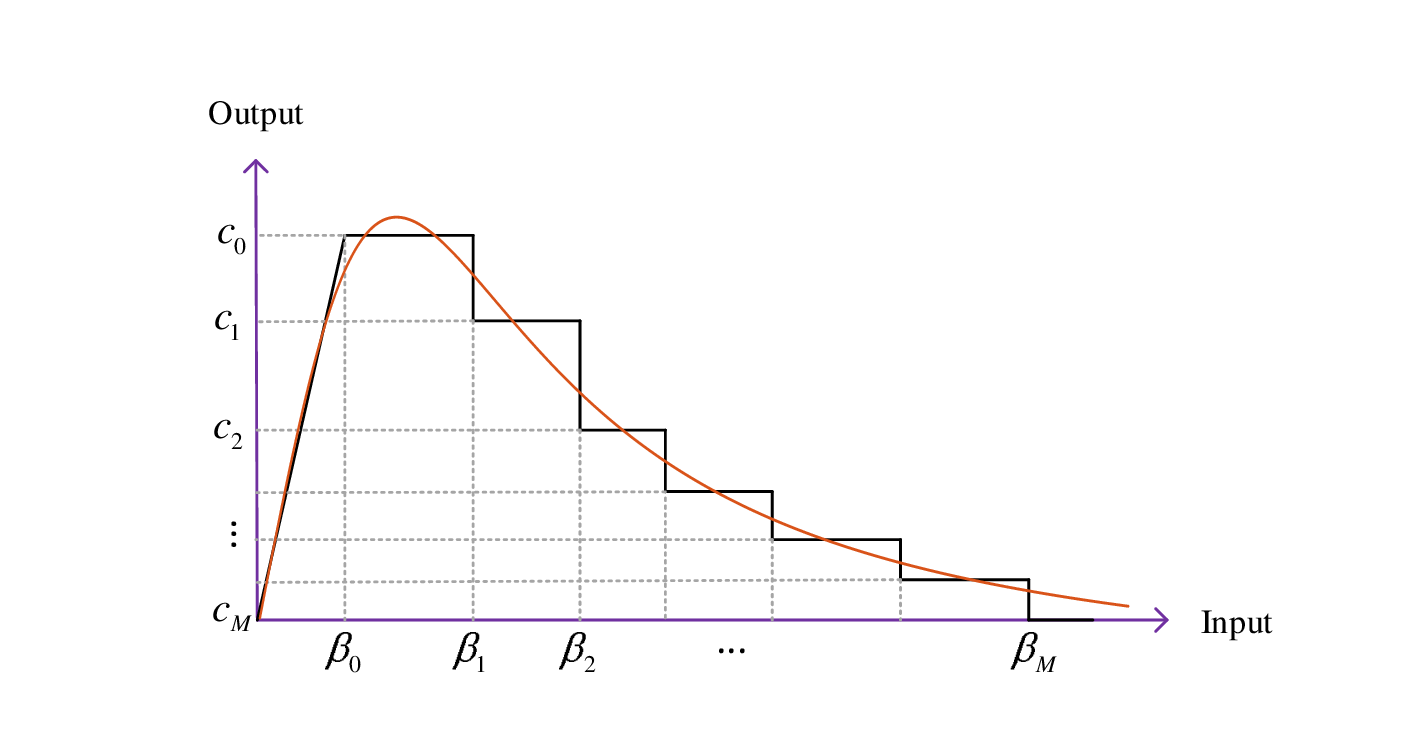}
\caption{Multi-threshold approximation of locally optimal detection.}
\label{fig:glo}
\vspace{-.6cm}
\end{figure}

Locally optimal detection of signals in non-Gaussian noise exploits nonlinear kernel~\cite{Introduction_VincentPoor_1998}. Based on the locally most powerful (LMP) test, for a given noise distribution, the optimal choice corresponds to
\begin{equation}
g(n_l) = -\frac{f^{'}_{n_l}}{f_{n_l}},
\end{equation}
where $f_{n_l}$ represents the probability density function of the interference amplitude and $f^{'}_{n_l}$ is its derivative. The exact shape of the optimum kernel may be too complicated to be implemented in practice~\cite{MultiThershold_Rozic_2018}. In addition, the interference parameters will change according to non-stationary nature of the dynamic environment. This non-stationary situation enforces the receiver to optimize the shape of the nonlinear function $g(.)$ for the optimal detection. In order to find the suboptimal shape of the nonlinear kernel, we investigate a multi-threshold clipper as shown in Fig.~\ref{fig:glo}. The proposed threshold-based clippers can be expressed as a linear combination of $M$ clippers:
\begin{equation}
g(r_k) = \sum\limits_{m = 0}^M {{c_m}\frac{{r_k}}{{\left| r_k \right|}}{u_m}\left( {\left| r_k \right|} \right)} ,
\end{equation}
where $r_k$ is the $k$-th sample of received OFDM symbol $\mathbf{r}$, $c_m$ is the clipping level in non-overlapping support on the $m$-th interval $[\beta_m,\beta_{m+1})$, and $u_m(.)$ is the unit box function given by
\begin{align}
{u_m}(r) = \left\{ \begin{gathered}
  1,\,\,\,\text{if}\,\,r_k \in \left[ \beta_m,\beta_{m + 1} \right), \hfill \\
  0,\,\,\,\text{otherwise}. \hfill \\
\end{gathered}  \right.
\end{align}
For the proposed multi-threshold clipper $c_0=\beta_0$, $c_M$ approaches zero and $\beta_M$ approaches $\max (\mathbf{r})$ to cover all the dynamic range of the incoming signal. In other words, as shown in Fig.~\ref{fig:glo}, the proposed multi-threshold clipper compromises between clipping and blanking in the presence of interference by cutting the incoming signal with different thresholds at different levels. It worths mentioning that the value of $\beta_0$ should ensure distortion-less processing of the incoming signal when there is no interference. In practice, the value of $M$ can be determined as a trade-off between performance of the system and complexity of the receiver.
\vspace{-.1em}
In general, finding the clipping levels $c_m$ and the threshold values $\beta_m$ is computationally expensive as both $c_m$ and $\beta_m$ are continuous variables. In order to reduce the complexity and find a suboptimal solution, we assume that $\beta_m$ can take a value in a predefined set $\mathcal B$ while satisfying following requirements
\begin{equation}\label{eq:Beta}
\begin{array}{l}
{\beta _m} \in \mathcal B,\,\, {\beta _m} < {\beta _{m + 1}},\\
{\beta _0} > \frac{1}{2}\max (\mathbf{x}_0),\,\,{\beta _M} \simeq \max (\mathbf{r}),
\end{array}
\end{equation}
where set $\mathcal B$ contains $n\times M$ different coefficients which can be selected uniformly from values between $\beta _0$ and $\beta _M$. In addition, $n$ is a constant which determines the quantization precision and can vary depending on the application. The same strategy can be invoked to find the suboptimal values for clipping levels $c_m$ such that
\begin{equation}\label{eq:C}
\begin{array}{l}
{c _m} \in \mathcal C,\,\, {c_m} \geq {c _{m + 1}},\,\, {c_0} = \beta_0,\,\,{c_M} \simeq 0,
\end{array}
\end{equation}
where set $\mathcal C$ includes $M$ different values which can be selected uniformly from values between $c_0$ and $c_M$. Therefore, our problem is to find $M$ optimum pairs $(\beta_m^*,c_m^*)$ such that
\begin{equation}\label{eq:Opt}
({\beta_m^*},{c_m^*})=\mathop {\arg \min }\limits_{({\beta _m},{c_m})} \,\text{BER}.
\end{equation}
In the following, we propose a learning based approach to solve the optimization problem in~\eqref{eq:Opt} subject to the constraints in~\eqref{eq:Beta} and~\eqref{eq:C} to determine the clipping levels $c_m$ and the threshold values $\beta_m$ for a given $M$.

\section{Proposed RL-Based Interference Mitigation Framework }\label{sec:RL}

In this section, we will develop a novel solution, based on MAB, to solve the proposed interference mitigation problem in~\eqref{eq:Opt}.

\subsection{Multi Armed Bandit: Preliminaries}\label{sec:MAB}

MAB is a class of sequential learning and decision-making problems in which an agent attempts to make an optimal decision within a stochastic environment and minimize its long-term regret~\cite{Auer_MAB_2002}. The so-called regret, can be defined as the expected total reward loss with respect to the optimal situation where the best decision is always taken. Since the regret distribution is unknown, the agent needs to explore each arm (action) to provide a good estimate of the expected regret from each arm to avoid converging to a local optimum action. In order to find the optimum actions we use \emph{decaying $\varepsilon$-greedy} policy where the value of $\varepsilon$ slowly decays over time. Assuming a network with one agent and finite number of $A$ arms in a set $\mathcal A$, pulling arm $a\in \mathcal A$ at time $t$ causes a random $R_a(t)$ regret for the agent. The average regret of an action $a$ after $J_a$ selections can be updated by
\begin{equation}\label{eq:Regret}
{Q_a}(t) = {Q_a}(t - 1) + \frac{1}{{{J_a}(t)}}\left( {{R_a}(t) - {Q_a}(t - 1)} \right),
\end{equation}
where ${Q_a}(t)$ and ${Q_a}(t-1)$ are the average regret of action $a$ at times $t$ and $t-1$, respectively. In fact, \eqref{eq:Regret} represents the \emph{incremental implementation} of sample average that requires to keep track of $Q_a(t)$ and $J_a(t)$ to compute the average regret for each action at its next occurrence. Pseudo code for action selection in $\varepsilon$-greedy policy based on \emph{incremental implementation} is shown in Algorithm~\ref{alg:MAB}.
\begin{algorithm}[t!]
\caption{: The proposed algorithm based on MAB}\label{alg:MAB}
\begin{algorithmic}[1]
    \For{\texttt{$a \in \mathcal A$}}\Comment(Initialization)

         $Q_a(0)\gets$ small number

         $J_a(0)\gets 0$
    \EndFor
    \State \textbf{Repeat forever:}
    \[{a^*} \leftarrow \left\{ \begin{array}{l}
\mathop {\arg \min }\limits_a {Q_a}(t) ,\,\,\,\,\,\ \text{with probability}\ 1 - \varepsilon \\
\text{random action} ,\,\,\,\,\,\ \text{with probability}\,\,\ \varepsilon
\end{array} \right.\]

$\text{BER}_{a^*}(t) \gets \text{determine regret for}\ a^*$

$J_{a^*}(t) \gets J_{a^*}(t-1)+1$

$Q_{a^*}(t) \gets Q_{a^*}(t-1)+\frac{1}{J_{a^*}(t)}[\text{BER}_{a^*}(t)-Q_{a^*}(t-1)]$

\end{algorithmic}
\vspace{-.1cm}
\end{algorithm}
In this algorithm, the value of $Q_a(0)$ for any action $a \in \mathcal A$ is initialized with a small number at the beginning which can also be used as a simple way to encourage exploration. This optimistic initialization ensures that all actions are tried several times and the system does a fair amount of exploration prior to the convergence.


\subsection{Proposed Interference Mitigation Framework as an MAB Problem}

In this problem, the receiver of $u_0$ acts as an agent and try to find the best clipping thresholds and levels at each time to minimize the BER as a regret. According to MAB formulation, any pair of $({\beta _m},{c_m})$ can be perceived as a two dimensional action and considering conditions in \eqref{eq:Beta} and~\eqref{eq:C}, the total number of different actions can be given by
\begin{equation}
L_{\mathcal A} = \overbrace{\frac{{(nM)!}}{{M!(nM - M)!}}}^{ \Im_M } \times {J_M},
\end{equation}
where $(.)!$ is factorial operation and  $\Im_M $ denotes the number of sets with length $M$ inside a set of length $n \times M$ and $J_M$ is the number of clipping level sets with length $M$. In order to find $J_M$ and the corresponding elements in each set, we can break down the problem by finding the number of paths (and trajectories) between top left to one of the most right elements (last column) in a matrix with the constraints that from each element you can either move only to right or diagonal to satisfy the constraint in~\eqref{eq:C}. The pseudo code for finding the value of $J_M$ and the corresponding trajectory is provided in Algorithm~\ref{alg:Path}. For example, Table~\ref{tab:Number of level} provides the value of $J_M$ for some $M$.
\begin{algorithm}[t]
\caption{: Clipping level sets with length $M$}\label{alg:Path}
\begin{algorithmic}[1]
\State \textbf{Initialization:}

$J_M \gets 0$, $\text{Path} \gets [\,]$

    \For{i = 1 to M}

         $D_i \gets$ Find all path from (1,1) to (i,M)

         $J_i\gets $ Number of path in $D_i$

         $J_M \gets J_M + J_i$

         $\text{Path} \gets$ Append $D_i$ to Path
    \EndFor

\end{algorithmic}
\end{algorithm}
\begin{table}[t]
\begin{center}
\captionsetup{labelfont=sc,labelsep=newline}
\caption{\sc{Number of clipping level sets}}
\begin{tabular}{ |c||c|c|c|c|c|c|c|c }
\hline
M & 1&2&3&4&5&6&7 \\
\hline
$J_M$ & 1&3&10&35&126&462&1716 \\
 \hline
\end{tabular}
\label{tab:Number of level}
\end{center}
\vspace{-.6cm}
\end{table}

After finding all possible action sets, one can invoke Algorithm~\ref{alg:MAB} to find the best action set in response to environment for minimizing the considered regret function which is BER in this work. In this framework, the number of possible actions increases exponentially with $M$ and $n$ (e.g., even for small values $M=5, n=2$, there are $31752$ different action sets) which increases the convergence time toward optimum action or even in decaying $\varepsilon$-greedy scenario the optimum action will be missed. To address this convergence issue, one can reduce the resolution by choosing $n=1$ for $M$ equally spaced sections and try to find the best value for $\beta_0$. In this case, $\beta_0$ can be given by
\begin{equation}
\beta_0 = \kappa \hat \beta_0,  \,\,\,\,\,\,\ 0.5 \leq \kappa \leq 10,
\end{equation}
where $\kappa$ is a correction coefficient and a course estimation of $\hat \beta_0$ can be found based on Neyman-Pearson criterion~\cite{AdaptiveNoiseMitigation-2010}. In practice, the range of $\kappa$ can be quantized to $q$ levels determined by the required performance and complexity. Following this simplification, the total number of different actions reduces to:
\begin{equation}
L_{\mathcal A} = q \times {J_M}.
\end{equation}
With this simplification, the convergence time of the proposed method significantly reduces. 

\section{Simulation Results}\label{sec:Simulation Results}

We consider a THz network composed of one serving BS, a target user, and $I$ interferers based on Poisson point process (PPP) with intensity $\lambda_I$ interferers per square meter. The OFDM symbols include 1024 subcariers and the bandwidth of the system is $1 GHz$. The fading channel is generated according to the model presented in \ref{sec:Channel Model} and the channel estimation is done after interference suppression by using pilot subcarriers which are equally spaced between subcarriers. The BER performance is used to compare the proposed MAB-based interference mitigation with two baseline approaches, namely, blanking and clipping. The threshold value for blanking and clipping is obtained based on the approach provided in \cite{AdaptiveNoiseMitigation-2010}. The serving BS knows the channel matrix $\textbf{H}_{0}$ for optimum beamforming and the AoA for interferers are sampled uniformly from $[-\pi/2,\pi/2]$ in every $1$ms.

In all simulation, we set the modulation constellation to QPSK, the energy of bit over noise is $E_b/N_0=0$ dB, $M=3$, $n=1$, the interval $0.5 \leq \kappa \leq 10$ is equally quantized by $q=20$ levels, and $\varepsilon$ decays from one with factor $1/(\text{Number of actions} \times 10)$ to ensure that all actions would be selected for enough number of times. As the desired signal and the interference will pass through different fading channels and the beamforming gain is different for each of them, the system performance is evaluated at different transmit signal to interference power ratios (SIR). Without loss of generality, it is assumed that the transmit power for all interferers are equal and they are using $N_t^I = 128$ antennas at the transmitter side.

\begin{figure}[t]
\centering
\includegraphics[width=.44\textwidth,height=50mm]{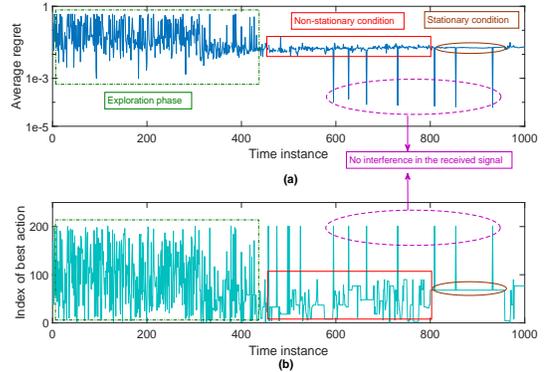}
\caption{(a) Average regret and (b) the corresponding selected action set over the time. $N_r^0=N_t^b=N_t^I=128$, $E_b/N_0$ =  0 dB, SIR = 0 dB, and $\lambda_I = 8\e^{-4}$.}
\label{fig:Cost_Action_MIMO}
\vspace{-.4cm}
\end{figure}

Fig.~\ref{fig:Cost_Action_MIMO} shows that the MAB-based approach converges to the optimum solution after evaluating all available action sets based on decaying $\varepsilon$-greedy algorithm. After converging to the optimum action set, the average regret in Fig.~\ref{fig:Cost_Action_MIMO}a will remain fixed unless the interference model or its parameters change in time. Therefore, the selected action set in Fig.~\ref{fig:Cost_Action_MIMO}b and its corresponding regret in Fig.~\ref{fig:Cost_Action_MIMO}a would always change unless it is in the stationary situation. Fig.~\ref{fig:Cost_Action_MIMO} demonstrates that when there is no interference in the received signal, the agent (receiver in UE $u_0$) would always select a specific action which do not harm the received signal.

\begin{figure}[t]
\centering
\includegraphics[width=.43\textwidth,height=52mm]{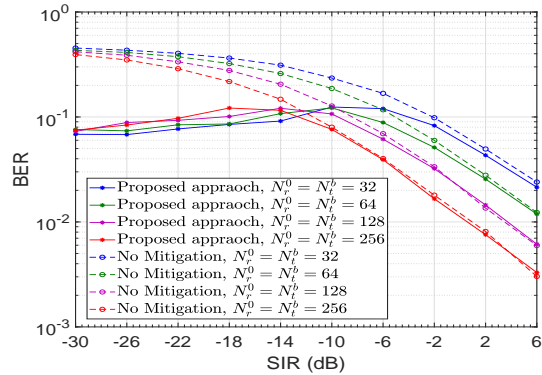}
\caption{BER performance of MAB-based interference mitigation for different number of antennas. $E_b/N_0$ =  0 dB, $\lambda_I = 8\e^{-4}$, and $N_t^I = 128$.}
\label{fig:MAB_SNR0_QPSK}
\vspace{-.4cm}
\end{figure}

Fig.~\ref{fig:MAB_SNR0_QPSK} illustrates the BER performance of the proposed MAB-based technique versus SIR for different number of antennas. As shown in Fig.~\ref{fig:MAB_SNR0_QPSK} the MAB technique improves the quality of the received signal specially at low SIRs. Since interference power is much smaller than the power of the desired signal at high SIRs, the interference is likely to be hidden in the received signal, and hence, it is difficult to detect and mitigate the interference by clipping the signal in the time domain. Thus, at high SIR (i.e., $\text{SIR} > -2 dB$) the performance of MAB would be the same as when there is no mitigation technique. However, the proposed MAB-based mitigation approach shows its potency by providing optimum thresholds and level values at low SIR (i.e., $SIR < -5 dB$) when the interference signals are distinguishable. This low SIR region is very important for THz communication as SIR is typically low if beam training is imperfect. According to Fig.~\ref{fig:MAB_SNR0_QPSK}, the performance of the MAB technique will slightly degrade as the SIR increases from -30 dB to -10 dB. At these SIR values, the amplitude of the interference starts to decay which makes it hard to distinguish the interference from the desired signal. Another interesting result from Fig.~\ref{fig:MAB_SNR0_QPSK} is that at low SIR region, having higher number of antennas at the transceivers of the desired link would not improve the performance of the proposed MAB-based interference reduction approach. This is due to the fact that by increasing the number of antennas and the corresponding beamforming gain, the received amplitude of the desired signal exceeds the amplitude of the interference signal and the MAB approach would not be able to detect the intermittent interference.

\begin{figure}[t]
\centering
\includegraphics[width=.42\textwidth,height=52mm]{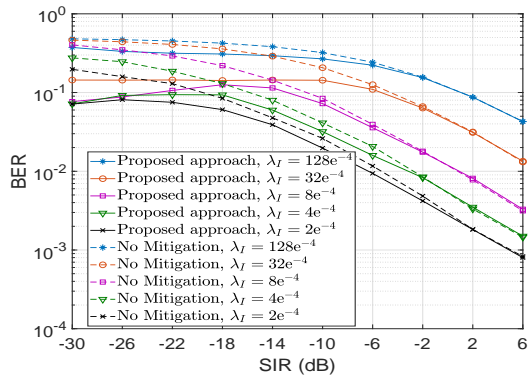}
\caption{BER performance of MAB-based interference mitigation for different density of interferer. $E_b/N_0$ =  0 dB, $N_r^0 = N_t^b = 256$, and $N_t^I = 128$.}
\label{fig:Density}
\vspace{-.4cm}
\end{figure}

Fig.~\ref{fig:Density} demonstrates the BER performance of proposed approach for different intensity of interferes versus SIR. It is clear from Fig.~\ref{fig:Density} that the BER performance will degrade as the number of interferes increases according to parameter $\lambda_I$. As expected, the MAB-based approach is more effective at low SIR region where interference has higher amplitude.

\begin{figure}[t]
\centering
\includegraphics[width=.43\textwidth,height=52mm]{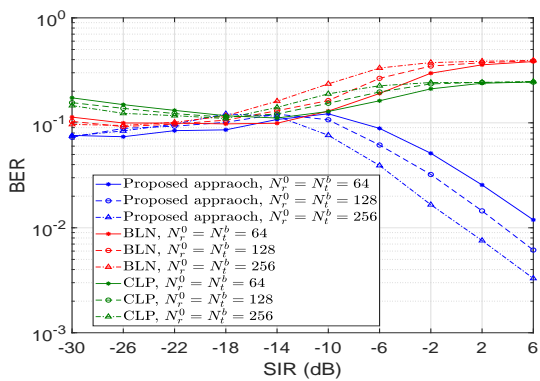}
\caption{Performance comparison between MAB, BLN, and CLP. $E_b/N_0$ =  0 dB, $\lambda_I = 8\e^{-4}$, and $N_t^I = 128$.}
\label{fig:BER_COMP_MAB}
\vspace{-.4cm}
\end{figure}

Fig.~\ref{fig:BER_COMP_MAB} compares the BER performance of the proposed MAB-based technique with blanking (BLN) and clipping (CLP) versus SIR for different numbers of antennas. As shown in Fig.~\ref{fig:BER_COMP_MAB}, the MAB-based approach outperforms both baseline methods in all scenarios. Finding the optimum threshold for BLN and CLP is very challenging at high SIRs as the level of peakedness decreases and it is difficult to find a proper threshold to distinguish between desired and contaminated signals. Thus, improper value for these threshold will corrupt the desired signal and significantly degrade the performance. Therefore, a fixed strategy for determining a single threshold for BLN and CLP will fail in many cases, especially in non-stationary scenarios. Although at very low SIR values (i.e., $\text{SIR} < -22 dB$) the performance of the proposed approach and BLN are close in some scenarios, the proposed scheme yields better BER in high SIR region. In fact, it is trivial to find the optimum threshold for BLN when the power of interference is much higher than desired signal and in this situation, discarding the received signal has better performance than clipping the signal.

\section{Conclusions}\label{sec:Conclusion}

In this paper, we have proposed a novel framework to mitigate intermittent interference, resulting from uncoordinated beam training transmissions, in dense wireless THz networks. In fact, we have developed a new adaptive multi-thresholding interference mitigation scheme which allows minimizing the non-stationary interference power in the time domain. To find the optimum thresholds, we have formulated the problem as a multi-armed bandit (MAB) framework with multiple thresholds and levels as a two-dimensional action set. The proposed MAB-based approach minimizes the BER as the regret function in the learning process and yields near-optimum values for multi-threshold clipping levels. The simulation results have shown that the proposed approach is fast-converging and can effectively reduce the search space. Results also have shown that the proposed MAB-based approach outperforms conventional methods such as blanking and single-threshold clipping schemes.

\vspace{-.15cm}

\bibliographystyle{IEEEtran}

\bibliography{IEEEabrv,Reference}

\begin{thebibliography}{10}
\providecommand{\url}[1]{#1}
\csname url@samestyle\endcsname
\providecommand{\newblock}{\relax}
\providecommand{\bibinfo}[2]{#2}
\providecommand{\BIBentrySTDinterwordspacing}{\spaceskip=0pt\relax}
\providecommand{\BIBentryALTinterwordstretchfactor}{4}
\providecommand{\BIBentryALTinterwordspacing}{\spaceskip=\fontdimen2\font plus
\BIBentryALTinterwordstretchfactor\fontdimen3\font minus
  \fontdimen4\font\relax}
\providecommand{\BIBforeignlanguage}[2]{{%
\expandafter\ifx\csname l@#1\endcsname\relax
\typeout{** WARNING: IEEEtran.bst: No hyphenation pattern has been}%
\typeout{** loaded for the language `#1'. Using the pattern for}%
\typeout{** the default language instead.}%
\else
\language=\csname l@#1\endcsname
\fi
#2}}
\providecommand{\BIBdecl}{\relax}
\BIBdecl

\bibitem{Omid_TWC_2017}
O.~{Semiari}, W.~{Saad}, and M.~{Bennis}, ``Joint millimeter wave and microwave
  resources allocation in cellular networks with dual-mode base stations,''
  \emph{IEEE Trans. Wireless Commun.}, vol.~16, no.~7, pp. 4802--4816, July
  2017.

\bibitem{Omid_WC_2019}
O.~{Semiari}, W.~{Saad}, M.~{Bennis}, and M.~{Debbah}, ``Integrated millimeter
  wave and {Sub-6 GHz} wireless networks a roadmap for joint mobile broadband
  and ultra-reliable low-latency communications,'' \emph{IEEE Wireless
  Commun.}, vol.~26, no.~2, pp. 109--115, April 2019.

\bibitem{Rappaport_TeraHz_2019}
S.~Ju \emph{et~al.}, ``Scattering mechanisms and modeling for terahertz
  wireless communications,'' in \emph{2019 IEEE Int. Commun. Conf.}, May 2019,
  pp. 1--7.

\bibitem{Rogalski_Terahertz_2011}
A.~Rogalski and F.~Sizov, ``Terahertz detectors and focal plane arrays,''
  \emph{Opto-Electronics Review}, vol.~19, no.~3, pp. 346--404, Sep 2011.

\bibitem{Cross2018niknam}
S.~Niknam, R.~Barazideh, and B.~Natarajan, ``Cross-layer interference modeling
  for 5{G} mmwave networks in the presence of blockage,'' in \emph{IEEE Veh.
  Tech. Conf.}, Aug. 2018, pp. 1--5.

\bibitem{Regime2018niknam}
S.~{Niknam} and B.~{Natarajan}, ``On the regimes in millimeter wave networks:
  Noise-limited or interference-limited?'' in \emph{IEEE Int. Conf. on Commun.
  Workshops}, May 2018, pp. 1--6.

\bibitem{Joham_MIMO_ZF_2005}
M.~{Joham}, W.~{Utschick}, and J.~A. {Nossek}, ``Linear transmit processing in
  {MIMO} communications systems,'' \emph{IEEE Trans. Sig. Process.}, vol.~53,
  no.~8, pp. 2700--2712, Aug 2005.

\bibitem{Wiesel_ZFBF_2008}
A.~{Wiesel}, Y.~C. {Eldar}, and S.~{Shamai}, ``Zero-forcing precoding and
  generalized inverses,'' \emph{IEEE Trans. Sig. Process.}, vol.~56, no.~9, pp.
  4409--4418, Sep. 2008.

\bibitem{Khodam_TVT}
R.~Barazideh, B.~Natarajan, A.~V. Nikitin, and S.~Niknam, ``Performance
  analysis of analog intermittently nonlinear filter in the presence of
  impulsive noise,'' \emph{{IEEE} Trans. Veh. Technol.}, vol.~68, no.~4, pp.
  3565--3573, Apr. 2019.

\bibitem{MultiThershold_Rozic_2018}
N.~Rozic, P.~Banelli, D.~Begusic, and J.~Radic, ``Multiple-threshold estimators
  for impulsive noise suppression in multicarrier communications,'' \emph{IEEE
  Trans. Signal Process.}, vol.~66, no.~6, pp. 1619--1633, Mar. 2018.

\bibitem{AdaptiveNoiseMitigation-2010}
G.~Ndo, P.~Siohan, and M.~H. Hamon, ``Adaptive noise mitigation in impulsive
  environment: Application to power-line communications,'' \emph{{IEEE} Trans.
  Power Del.}, vol.~25, no.~2, pp. 647--656, Apr. 2010.

\bibitem{Khodam_CCWC_ML}
R.~Barazideh, S.~Niknam, and B.~Natarajan, ``Impulsive noise detection in
  {OFDM}-based system: A deep learning perspective,'' in \emph{IEEE Comput.
  Commun. Work. Conf. (CCWC)}, Jan. 2019, pp. 0937--0942.

\bibitem{Akdeniz_mmWaveChannelModel_2014}
M.~R. {Akdeniz}, Y.~{Liu}, M.~K. {Samimi}, S.~{Sun}, S.~{Rangan}, T.~S.
  {Rappaport}, and E.~{Erkip}, ``Millimeter wave channel modeling and cellular
  capacity evaluation,'' \emph{IEEE Journal on Selected Areas in
  Communications}, vol.~32, no.~6, pp. 1164--1179, Jun. 2014.

\bibitem{VA_BeamwidthandChVar_2017}
V.~{Va}, J.~{Choi}, and R.~W. {Heath}, ``The impact of beamwidth on temporal
  channel variation in vehicular channels and its implications,'' \emph{IEEE
  Trans. Vehi. Technol.}, vol.~66, no.~6, pp. 5014--5029, June 2017.

\bibitem{Introduction_VincentPoor_1998}
H.~V. Poor, \emph{An introduction to signal detection and estimation (2nd
  ed.)}.\hskip 1em plus 0.5em minus 0.4em\relax Springer, 1994.

\bibitem{Auer_MAB_2002}
P.~Auer, N.~Cesa-Bianchi, and P.~Fischer, ``Finite-time analysis of the
  multiarmed bandit problem,'' \emph{Machine Learning}, vol.~47, no.~2, pp.
  235--256, May 2002.

\end{thebibliography}

\end{document}